\documentclass[twocolumn,showpacs,nofootinbib]{revtex4}

\include{amssymbol}
\input{epsf}

\def\mpc{\mbox{Mpc}}

\def\AJ{{\it Ap. J.} }
\def\AJL{{\it Ap. J. Lett.} }

\def\ANJ{{\it Astron. J.} }

\def\ASAS{{\it Astron. and Astrophys.} }

\def\BAPS{{\it Bull. Am. Phys. Soc.} }

\def\CQG{{\it Class. Quantum Gravity} }

\def\GRG{{\it Gen. Relativity and Gravitation} }

\def\IJMP{{\it Int. J. Mod. Phys.} }

\def\MNRAS{{\it Mon. Not. R. Ast. Soc.} }

\def\NC{{\it Il Nuovo Cimento} }
\def\NP{{\it Nucl. Phys.} }
\def\PL{{\it Phys. Lett.} }
\def\PR{{\it Phys. Rev.} }
\def\PRL{{\it Phys. Rev. Lett.} }

\def\RMP{{\it Rev. Mod. Phys.} }

\def\al{\alpha} \def\be{\beta}  
\def\ep{\epsilon}   
\def\th{\theta}   \def\ka{\kappa}
\def\la{\lambda}   
\def\si{\sigma}   
   
\def\om{\omega} \def\Ga{\Gamma} \def\De{\Delta} 
\def\La{\Lambda}   
 \def\Om{\Omega} \def\mn{{\mu\nu}} \def\cl{{\cal L}}

 \def\frac#1#2{{\textstyle{{#1}\over
{#2}}}} 
\def\lsim{\mathrel{\rlap{\lower4pt\hbox{\hskip1pt$\sim$}}
\raise1pt\hbox{$<$}}}
\def\gsim{\mathrel{\rlap{\lower4pt\hbox{\hskip1pt$\sim$}}
\raise1pt\hbox{$>$}}} \def\sqr#1#2{{\vcenter{\vbox{\hrule height.#2pt
\hbox{\vrule width.#2pt height#1pt \kern#1pt \vrule width.#2pt} \hrule
height.#2pt}}}}
\def\square{\mathchoice\sqr66\sqr66\sqr{2.1}3\sqr{1.5}3}
\def\beq{\begin{equation}} \def\eeq{\end{equation}}
\def\beqa{\begin{eqnarray}} \def\eeqa{\end{eqnarray}}
\def\eq#1{Eq. (\ref{#1})}

\begin{document}

\title{Mimicking the cosmological constant: \\constant curvature spherical solutions in a non-minimally coupled model}

\author{Orfeu Bertolami\footnote{Also at Instituto de Plasmas e F\'isica Nuclear, Instituto Superior T\'ecnico, Av. Rovisco Pais, 1, 1049-001, Lisboa Portugal.}}
\email{orfeu.bertolami@fc.up.pt}
\affiliation{Departamento de F\'{\i}sica e Astronomia, Faculdade de Ci\^encias, Universidade do Porto,\\Rua do Campo Alegre 687,
4169-007 Porto, Portugal}

\author{Jorge P\'aramos}
\email{paramos@ist.edu}
\affiliation{Instituto de Plasmas e Fus\~ao Nuclear, Instituto Superior T\'ecnico\\Av. Rovisco Pais 1, 1049-001 Lisboa, Portugal}

\date{\today}

\begin{abstract}
The purpose of this study is to describe a perfect fluid matter distribution that leads to a constant curvature region, thanks to the effect of a non-minimal coupling. This distribution exhibits a density profile within the range found in the interstellar medium and an adequate matching of the metric components at its boundary. By identifying this constant curvature with the value of the cosmological constant, and superimposing the spherical distributions arising from different matter sources throughout the universe, one is able to mimic a large-scale homogeneous cosmological constant solution.

\end{abstract}


\maketitle 

\section{Introduction}

Contemporary cosmology faces currently major challenges. These include, for instance, the need to firmly establish the existence and nature of dark energy and dark matter, or to find evidence of a more encompassing theory of gravity. The nature of dark energy \cite{DE} has been discussed in several proposals, including quintessence models, which consider the slow-roll of a scalar field \cite{quintessence,Copeland}, averaging of inhomogeneities at a cosmological scale yielding an effective scalar field \cite{inhomo}, amongst other candidates. A possible unification of both ``dark'' components has also been suggested, resorting to a scalar field model \cite{Rosenfeld} or an exotic equation of state, as featured by the so-called generalized Chaplygin gas \cite{Chaplygin}.

Another outstanding difficulty in the context of the General Relativity (GR) description of the universe dynamics and its relationship with the standard model of particle physics is the cosmological constant problem, a notorious adjustment issue that so far lacks a proper explanation (see {\it e.g.} \cite{OB2009} and references therein for a discussion and a list of the most well known proposals for a solution). A different approach assumes that no extra energy content is needed, but that the fundamental laws and tenets of gravitation are incomplete, such that with current knowledge just offers a low-energy perspective of a more fundamental theory; as a consequence, modifications of the Friedmann equation to include higher order terms in the energy density $\rho$ (see {\it e.g.} \cite{Maartens,scalar} and references therein) have been proposed or, at a more fundamental level, changes to the action functional. A rather straightforward approach lies in replacing the linear scal
 ar curvature term in the Einstein-Hilbert action with a function of the scalar curvature, $f(R)$. Alternatively, one could resort to other scalar invariants of the theory \cite{f(R)}: extensions relying on a functional dependence of the action on the Gauss-Bonnet invariant $G= R^2 - 4R_\mn R^\mn + R_{\al\be\mn} R^{\al\be\mn}$ are perhaps the most well studied models, as this could arise in a low-energy effective description of String Theory and carry several implications in braneworld scenarios  \cite{GB}. The  more tractable class of $f(R)$ theories have had considerable success in replicating the early period of rapid expansion of the universe, with the Starobinsky inflationary model $f(R)=R + \al R^2$ being the most well-known proposal \cite{Staro}. The accelerated expansion of the universe has also been the subject of attention \cite{capoexp}. Solar system tests could also bring further insight, mostly arising from the parameterized post-Newtonian (PPN) metric coefficien
 ts derived from this extension of GR. However, some disagreement exists in the community, with some arguing that no changes are predicted at a post-Newtonian level (see {\textit e.g.} \cite{PPN} and references therein); amongst other considerations, this mostly stems from an approach based either in the more usual metric affine connection (that is, where the affine connection is taken {\it a priori} as depending on the metric), or in the so-called Palatini approach \cite{Palatini} (where both the metric and the affine connection are taken as independent variables). As an example of a clear phenomenological consequence of this extension of GR, it has been shown that $f(R) = f_0 R^n$ theories yield a gravitational potential which displays an increasing, repulsive contribution, added to the Newtonian term \cite{flat}.

Notwithstanding the significant literature on these $f(R)$ models, steps have been taken to explore another interesting possibility: not only that the curvature is non-trivial in the Einstein-Hilbert Lagrangian, but also that the coupling between matter and geometry is non-minimal, {\it i.e.} enforced solely by the invariant $\sqrt{-g} d^4x$, the use of the metric to raise and lower indexes and the associated covariant derivative; indeed, in GR covariantly invariant terms in $\mathcal{L}_m$ should be constructed by contraction with the metric ({\it e.g.} the kinetic term of a real scalar field, $g^\mn \phi_{,\mu} \phi_{,\nu}$): a non-minimum coupling would imply that geometric quantities (such as the scalar invariants) would explicitly appear in the action. This possibility can have deep phenomenological implications: in particular, it would imply that regions with high curvature (which, in GR, are related to regions of high energy density or pressure) could lead to considera
 ble deviations from the dynamics predicted by Einstein's theory \cite{Lobo}. A wide range of results has arisen from this hypothesis, including the formal equivalence with a multi-scalar-tensor theory \cite{equivalence}, the impact on solar observables \cite{solar}, the possibility to account for dark matter \cite{DM}, post-inflationary reheating \cite{reheating} or the current accelerated expansion of the universe \cite{cosmo}.

When studying the latter, it was noted that one cannot recover an exponential expansion due to a constant curvature, {\it i.e.} a universe dominated by a cosmological constant (CC), when working under the assumption of a Friedmann-Robertson-Walker metric. Since this is the simplest of all culprits behind the mentioned accelerated expansion, one now reassesses this issue and attempts to obtain a local, spherically symmetric solution for the interstellar matter (ISM) distribution that, in the context of a non-minimally coupled model, gives rise to a constant scalar curvature. This yields a mechanism that generates what can be viewed as a local CC. The superposition of several spherical distributions throughout the universe could then lead to a universal CC. Of course, this is not a solution for the cosmological constant problem, but a proposal to account for the observations once a mechanism is found to make all other contributions vanish.

The goal of this work is then to achieve this description without the requirement of a specific model for the non-minimal coupling or a somewhat awkward and convoluted density profile --- {\it i.e.} although one shows that only a quite specific distribution gives rise to a constant curvature, this can be regarded as an admissible profile for the ISM medium. One may also regard this work as yet another example of the strong impact a non-minimal coupling can have in the description of matter distribution, when compared with the standard results arising from GR.

Related proposals aiming to describe, for instance, dark energy through some variation of the concept of a non-minimal coupling between matter and geometry have been put forward previously, addressing the problem of the accelerated expansion of the universe \cite{expansion} and the existence of a CC \cite{cosmological}. However, one posits that the approach proposed here is somewhat more economical and elegant, as it does not explicitly introduce any additional matter field, and it does not set any specific form for the non-minimally coupled terms.

\section{The model}

One considers a model that exhibits a non-minimal coupling between geometry and matter, as expressed in the action functional \cite{Lobo},

\beq S = \int \left[ \ka R + f_2(R) \mathcal{L} \right] \sqrt{-g} d^4 x~~. \label{model}\eeq

Variation with respect to the action yields the modified Einstein field equations,

\beqa && 2\left(\ka +  F_2 \cl \right) \left(R_\mn - {1 \over 2}g_\mn R\right) = \\ \nonumber && f_2 T_\mn - F_2 \cl R g_\mn +2 \De_\mn \left(F_2 \cl\right)~~, \label{field0}\eeqa

\noindent with $\ka = c^4/16\pi G$, $\De_\mn = \nabla_\mu \nabla_\nu - g_\mn \square$ and $\cl = -\rho$ (see Ref. \cite{fluid} for a discussion about this choice within the context of a non-minimal coupled model). As expected, GR is recovered by setting $f_1(R) = 2\ka R $ and $f_2(R) = 1$.

Resorting to the Bianchi identities, one concludes that the energy-momentum tensor of matter may not be conserved in a covariant way, as

\beq \nabla_\mu T^\mn={F_2 \over f_2}\left(g^\mn \cl-T^\mn\right)\nabla_\mu R~~. \label{cov} \eeq

\noindent Again, in the absence of a non-minimal coupling, $f_2(R)=1$ and the covariant conservation of the energy-momentum tensor is recovered. This feature implies that the motion of the matter distribution described by a Lagrangian density $\cl$ does not follow a geodesic curve. Clearly, a violation of the Equivalence Principle may emerge if the {\it r.h.s.} of the last equation varies significantly for different matter distributions, which suggests a method of testing the model and imposing constraints on the associated couplings.

A constant curvature leads to $R_\mn = R_0 g_\mn/4 \rightarrow R = R_0$ and $\cl = -\rho$; inserting this into \eq{field0}, one gets

\beq \left(\ka + F_2 \rho \right)g_\mn R_0 = 4 F_2 \De_\mn\rho - 2 f_2 T_\mn  ~~. \label{field1}\eeq

One writes the metric as 

\beq \label{metric} ds^2 = - e^{2\nu(r)} dt^2 + e^{2\si(r)} dr^2 + d\Om^2~~. \eeq

\noindent Assuming a static configuration, $u_\mu = (u_0,\vec{0})$; since $u_\mu u^\mu = g^{00} u_0^2 = -1$, one gets $u_0 = (-g_{00})^{1/2} = e^{\nu(r)}$. Thus, the perfect fluid form for the energy-momentum tensor 

\beq T_\mn = (\rho + p)u_\mu u_\nu + p g_\mn \rightarrow T = 3p - \rho~~, \eeq

\noindent reads

\beq T_{00} = \rho e^{2\nu}~~~~,~~~~ T_{rr} = pe^{2\si} ~~~~,~~~~T_{\th\th} = pr^2~~. \eeq

One assumes a constant equation of state (EOS) parameter $0 \lesssim\om = p/\rho \leq 1$, where $\om = 0$, $1/3$ and $1$ gives dust, ultra-relativistic and ultra-stiff matter, respectively. Taking the trace, one obtains

\beq \left(\ka + F_2 \rho \right) R_0 = -3 F_2 \square \rho + {1 \over 2} f_2 (1 - 3\om)\rho  ~~. \label{trace}\eeq

\noindent Replacing into \eq{field1}, one gets

\beq  \left[  {1 \over 2} f_2 (1 - 3\om) -3 F_2 \square \right] g_\mn \rho = 4 F_2 \De_\mn\rho - 2 f_2 T_\mn  ~~. \label{field2}\eeq

One then has

\beqa \nabla_\mu \nabla_\nu \rho &=& \nabla_\mu \rho_{,\nu} = \rho_{,\mn} - \Ga_{\mn}^\al \rho_{,\al} \rightarrow \\ \nonumber \nabla_0 \nabla_0 \rho &=&  - \Ga_{00}^r \rho' =  {1 \over 2} g^{rr} g_{00,r} \rho' = - e^{2(\nu-\si)} \nu' \rho' ~~,\\ \nonumber  \nabla_r \nabla_r \rho &=& \rho'' - \Ga_{rr}^r \rho' = \rho'' -\si' \rho' ~~, \\ \nonumber \nabla_\th \nabla_\th \rho &=&  - \Ga_{\th\th}^r \rho' = {1 \over 2}g^{rr} g_{\th\th,r}\rho' = e^{-2\si} r\rho' ~~, \eeqa

\noindent and

\beqa e^{2\si}\square \rho &=& {e^{2\si} \over \sqrt{-g}} \left(g^{\mn} \sqrt{-g} \rho_{,\nu} \right)_{,\mu} = \\ \nonumber && {1 \over e^{\nu-\si} r^2 \sin \th } \left(e^{\nu-\si} r^2 \sin \th \rho'\right)' = \\ \nonumber && \rho'' + {2 \over r} \rho'  + (\nu'-\si')\rho' ~~,\eeqa

\noindent so that

\beqa \De_{00} \rho &=& e^{2(\nu-\si)} \left[ \rho'' + \left({2 \over r}-\si' \right)\rho' \right] ~~, \\ \nonumber \De_{rr} \rho &=& - \left({2 \over r} + \nu'  \right) \rho'~~, \\ \nonumber \De_{\th\th} \rho &=& -e^{-2\si}r^2 \left[\rho'' + \left( {1 \over r} + \nu' - \si' \right) \rho' \right] ~~, \eeqa

\noindent where $\rho' \equiv \rho_{,r}$ and one assumes only a radial dependency for all quantities.

Since $e^{2\si}g_{00}+ e^{2\nu}g_{rr} = 0$, one obtains from \eq{field2}

\beq \rho'' - \left(\nu' + \si'\right)\rho' = {1 +\om \over 2}{f_2 \over F_2} e^{2\si} \rho ~~, \label{combo} \eeq

\noindent while the $(\th,\th)$ component of the field \eq{field2} reads

\beq -\rho'' + \left( {2 \over r} - \nu' + \si' \right) \rho' = {1 +\om \over 2}{f_2 \over F_2} e^{2\si} \rho ~~. \label{fieldthth}\eeq

\noindent Notice that both equations show that the only constant solution is $\rho = 0 \rightarrow R_0 = 0$, as in GR (unless $f_2(R_0) = 0$, which is unphysical). If $F_2 \neq 0$, equating the two equations above yields

\beq \label{exact}\rho'' = \left( {1 \over r} + \si'  \right) \rho' \rightarrow \rho' = K R_0 e^\si r ~~,\eeq

\noindent where $K$ is a constant with dimensions of density. Notice that $K=0$ leads to $\rho =  0 $ or $\om = -1$, the GR conditions for constant curvature.

Adding Eqs. (\ref{combo}) and (\ref{fieldthth}) yields

\beq f_2 (1 +\om)\rho e^{2\si} = 2F_2 \left(  {1 \over r} - \nu' \right) \rho' ~~, \eeq

\noindent which, again using \eq{combo}, reads

\beq (\rho^2)' = {4(K R_0)^2\over 1+\om} {F_2 \over f_2} r \left(  1 - \nu' r \right)~~, \eeq

\noindent implicitly defining $\rho$ in terms of $\nu$.

In order to obtain a single differential equation for $\rho$, one first isolates the metric coefficients as

\beqa \label{metricK} e^{2\si} &=& \left( {\rho' \over KR_0r}\right)^2 ~~, \\ \nonumber \si' &=& {\rho'' \over \rho'}-{1 \over r}~~,\\ \nonumber \nu' &=&  {1 \over r} - {1+\om \over 2}{f_2 \over F_2(KR_0r)^2} \rho \rho' \rightarrow \\ \nonumber \nu'' & = & -{1 \over r^2} - {1+\om \over 2(KR_0r)^2}{f_2 \over F_2} \left[ (\rho')^2 +  \rho \rho'' - {2 \over r} \rho \rho' \right] ~~. \eeqa

One now takes the definition of the scalar curvature,

\beqa  {e^{2\si} \over 2} R = {e^{2\si} - 1 \over r^2} + \left({2 \over r} + \nu '\right) \left(\si' -\nu ' \right) - \nu '' ~~, \eeqa

\noindent and, replacing \eq{metricK}, obtains a differential equation for $\rho$,

\beqa  \label{difeq} && \rho'' - {2 \over r} \rho' + {1+\om \over 2(KR_0r)^2}{f_2 \over F_2} \rho (\rho')^2 \\ \nonumber && - {1\over 3r^3} \left({ 1 + \om \over 2(KR_0)^2}\right)^2 \left({f_2 \over F_2}\right)^2  \rho^2 (\rho')^3 + \\ \nonumber && + {1 \over 6(KR_0)^2r} (\rho')^3 \left( {2 \over r^2} - R_0 + {1 + \om \over 2}{f_2 \over F_2} \right) = 0~~.\eeqa

In the case of the minimal coupling $f_2(R) = 1$, this is ill-defined unless $\rho' = 0$, as discussed before. This is also attained by setting $K=0$.

\subsection{Boundary matching conditions}

The purpose of this work is to show that one may locally mimic a CC via a suitable matter distribution, in the context of a non-minimal coupling between the latter and curvature. In line with the Cosmological Principle, it was argued in the introduction that, even in a cosmological context, $\La$ arises not as a fundamental constant of nature, but from an averaging of all the surrounding matter distributions with constant curvature (perhaps even each with a different curvature). A more careful study of the averaging of  matter distributions with different values of the (constant) curvature is delayed for a future work (see {\it e.g.} \cite{average}, as well as \cite{average2} for a discussion of the Einstein-Strauss and Lindquist-Wheeler solutions for the issue of embedding a local ``bubble'' in an expanding spacetime issue in GR).

However, one cannot simply assume that such averaging yields a constant density space surrounding the interior solution under scrutiny: from Eqs. (\ref{combo}) and (\ref{fieldthth}), one sees that the only constant solution is $\rho = 0$: this is expected, as this work shows that a constant curvature enforces a specific non-flat profile for the density.

In order to circumvent this issue, one assumes that the interior solution is surrounded instead by an exterior vacuum solution with a constant curvature $R = 4\La$. In this region, $\rho = 0$ and the effect of the non-minimal coupling vanishes: one recovers the usual Schwarzschild-de Sitter metric,

\beq e^{-2\si} = e^{2\nu} = 1 - {M \over 8\pi \ka r} -\La r^2~~, \eeq

\noindent so that

\beq \si'=-\nu' = { \La r - {M\over 16\pi\ka r^2} \over 1 - {M\over 8\pi \ka r} - \La r^2} ~~. \eeq

In order to match the interior and exterior solutions at the radius of the spherical body $r=r_s$, defined as $\rho(r_s)=0$, \eq{metricK} lead to

\beqa \label{matching} {\rho''(r_s) \over \rho'(r_s)}-{1 \over r_s} = { \La r_s - {M\over 16 \pi \ka r_s^2} \over 1 - {M\over 8\pi\ka r_s} - \La r_s^2} ~~, \\ {1 \over r_s} = -{ \La r_s - {M\over 16 \pi \ka r_s^2} \over 1 - {M\over 8\pi \ka r_s} - \La r_s^2} ~~,   \eeqa

\noindent thus implying that $ \rho''(r_s) = 0 $; this does not depend upon the specific form of $\nu$ and $\si$, but only on the boundary conditions $\nu'(r_s) = -\si'(r_s)$ and $\rho(r_s) = 0$. Indeed, inserting these into \eq{combo} leads to 

\beq 2F_2\rho''(r_s) = e^{2\si(r_s)} f_2(1+\om) \rho(r_s) = 0~. \eeq

\noindent Inserting $\rho''(r_s) = 0 $ into \eq{matching}, one obtains a relation between the total mass $M$ and $r_s$, independent of $\La$ and other parameters:

\beq M={16\pi \over 3} \ka r_s ~~.  \eeq

\noindent Notice that $M$ is not necessarily the gravitational mass, but merely an integration constant of the field equations in vacuum.

Using \eq{exact} to evaluate \eq{difeq} at $r=r_s$, one obtains 

\beq 12(KR_0)^2 = { 3(KR_0)^2 r^2_s \over 1 - 3\La r_s^2}\left( {2 \over r_s^2} - R_0 + {1 + \om \over 2}{f_2 \over F_2} \right) ~~, \eeq

\noindent yielding

\beq \label{radius} r_s = \left( {1 + \om \over 4}{f_2 \over F_2} +6 \La -{R_0 \over 2} \right)^{-1/2} ~~. \eeq

Continuity of $\nu$ is imposed by an appropriate integration constant, since only its derivatives show in the dynamics. Regarding $\si$, matching at $r=r_s$ shows that the constant $K$ is related with the slope of the density at the boundary of the spherical body,

\beqa&& 1 - {M \over 8\pi \ka r_s} -\La r_s^2 = \left( {\rho'(r_s) \over KR_0r_s}\right)^{-2} \rightarrow \\ \nonumber && K = {\rho'(r_s) \over R_0} \sqrt{{1  \over 3r_s^2} -\La }~~. \eeqa

\noindent which is also obtained by evaluating \eq{exact} at $r=r_s$.

\section{Numerical analysis}

In order to pursue a numerical analysis of the suggested scenario, one introduces the dimensionless scaling

\beqa  \label{scaling} \rho = -\ep\sqrt{2 \over 1 + 3 \la - \ep } K \th ~~~~&,&~~~~ x = {r \over r_s} ~~, \\ \nonumber \ep = {2 \over 1+\om} {F_2 R_0 \over f_2 }~~~~&,& ~~~~\la = {8 \over 1+\om}{F_2 \La \over f_2}~~, \eeqa

\noindent so that $0<x<1$ and 

\beq \label{rs} r_s = \sqrt{\la \over 1 + 3 \la - \ep } {d_H \over \sqrt{6 \Om_\La }} ~~,\eeq

\noindent having used 

\beq \label{cc} \La = {8 \pi G \over c^2} \Om_\La \rho_c = 3 \Om_\La \left({H \over c}\right)^2 = {3 \Om_\La \over d_H^2}~~, \eeq

\noindent with $d_H = c/H_0 = 4.22~Gpc$ the Hubble distance.

Notice that a negative sign is used in the first definition of \eq{scaling}, so that a decreasing density $\rho' < 0 \rightarrow K < 0$ ({\it cf.} \eq{exact}) yields a positive $\th(x)$ function. With this transformation, \eq{difeq} only involves the parameters $\ep$ and $\la$,

\beqa  \label{difeqx} && \th'' - {2 \over x} \th' + {1 \over  x^2} \th (\th')^2 - {1 \over 3  x^3}\th^2 (\th')^3 + \\ \nonumber && {1 \over 6 x } (\th')^3 \left( {1+3\la - \ep \over x^2} - \ep + 1 \right) = 0~~,\eeqa

\noindent where the primes now indicate differentiation with respect to $x$.

\noindent The boundary conditions for the density translate into 

\beq \label{boundaryth} \th(1) = 0 ~~~~,~~~~\th'(1) = -2\sqrt{3 \over 2 +3 \la - 2\ep }  ~~, \eeq

\noindent as can be checked by evaluating \eq{difeqx} at $x=1$ and taking $\th''(1)=\rho''(r_s)=0$.

The numerical procedure is simple: for a given $K$, $R_0$, $\om$ and $f_2$, compute $\ep$ and $\la$. Evolve the above differential equation between $x=0$ and $x=1$, with the boundary conditions conditions above, and then read the density from \eq{scaling}. Notice that the constant $K$ does not appear in \eq{difeqx}; one may view it as a scaling parameter set by the central density of the spherical matter distribution,

\beq \rho_c = -K \th(0) \sqrt{2 \over 1 + 3\la - \ep } \ep >0~~, \eeq

\noindent since one expects the density to drop and $K< 0$.

\subsection{Smooth transition}

Before continuing, one addresses another possibility: instead of imposing continuity with an outer Schwarzschild-de Sitter metric, one could perhaps define the boundary of the spherical mass distribution as a radius $r=r_{out}$ where $\rho(r_{out})=\rho_{out}$ and $\rho'(r_{out})=0$, {\it i.e.} where a smooth transition to an exterior solution occurs. Notice that $r_{out} \neq r_s$, the latter being defined through \eq{radius} and characterizing the interplay between the model parameters. However, this cannot be done, as it can be seen from \eq{exact} that $\rho'(r_{out})=0 \rightarrow r_{out}) = 0$ or $K = 0$; the latter yields a constant $\rho = \rho_{out}$ sphere with no physically meaningful boundary with the identical distribution surrounding it, and which only allows for a vanishing constant curvature $R=0$, as discussed before.

\subsection{Metric elements}

One may rewrite \eq{metricK} in terms of the redefined quantities given by \eq{scaling},

\beqa \label{metricscaling} e^{2\si} &=& \left( {\rho' \over KR_0r}\right)^2 = { 1 + 3 \la - \ep \over 2 }\left[ { \th'(x) \over   x}\right]^2 ~~,\\ \nonumber \nu'(x) &=& r_s\left[{1 \over r} - {1+\om \over 2}{f_2 \rho \rho'\over F_2(KR_0r)^2}  \right]= {1 \over x}  - {\th(x) \th'(x) \over x^2} ~~. \eeqa

\noindent As discussed, these do not depend upon the integration constant $K$, showing that the direct physical meaning of $\th$ and its derivative is in the metric itself.

As discussed above, the boundary conditions and their scaling may be extracted from the matching with the rescaled Schwarzschild-de Sitter metric, after substituting $M=(16\pi / 3) \ka r_s$, which reads

\beqa \label{SdSscaling} e^{-2\si}&=&e^{2\nu} = 1 - {2 \over 3 x} - {\la \over 2(1 + 3\la - \ep)} x^2~~, \\ \nonumber \si'(x)&=&-\nu'(x) = {1 \over x}{2(1+3\la-\ep) - 3 \la x^3 \over 2(2-3x)(1+3\la - \ep)+3 \la x^3} ~~, \eeqa

\noindent By matching the two sets of equations, one again recovers the boundary conditions \eq{boundaryth}, as expected.

From \eq{SdSscaling}, one has

\beq e^{2\nu(1)} = {1 \over 3} - {\la \over 2(1 + 3\la - \ep)}~~.\eeq

\noindent Integrating the first of \eq{metricscaling}, one finds that

\beqa e^{2\nu(x)} &=& \left[{1 \over 3} - {\la \over 2(1 + 3\la - \ep)} \right] x^2 \times \\ \nonumber && \exp \left[ 2\int_x^1 {1 \over x^2}\th(x) \th'(x) dx \right]~~,\eeqa

\noindent  setting the integration constant that provides the matching of this metric element at the boundary $r= r_s$, as was discussed following \eq{radius}. Assuming a smooth $\th(x)$, one can see that no singularity arises, since inspection of the above expressions shows that neither $e^{2\si}$ nor $e^{2\nu}$ vanish.

\subsection{Outer Schwarzschild-de Sitter metric}

As discussed before, the mass parameter in the outer Schwarzschild-de Sitter metric does not need to reflect the gravitational mass of the perfect fluid matter distribution, as it is an integration constant of the Einstein equations in vacuum. As a result, one has shown that continuity of the inner and outer metric requires that $M= (16 \pi/3)\ka r_s $.

The gravitational mass $M_g$ is defined as

\beq M_g = 4\pi \int_0^{r_s} \rho r^2 dr = 4\pi \rho_c {r_s^3 \over \th(0)} \int_0^1 \th(x) x^2 dx~~.\eeq

\noindent If one aims to circumvent this distinction between $M$ and the total mass of the spherical body, the condition $M_g = M $ yields

\beq r_s = 2\sqrt{\ka \th(0) \over 3 \rho_c I}~~,\eeq

\noindent with 

\beq I = \int_0^1 \th(x) x^2 dx ~~.\eeq

Resorting to \eq{rs}, one finds that this sets the central density of the spherical body as

\beq \label{rscondition} \rho_c  = 8 \ka\left( { 1 + 3 \la - \ep \over \la} \right){\Om_\La \over d^2_H } {\th(0) \over I } ~~. \eeq

\section{Mimicking a cosmological constant}

One now searches for a solution that mimics a cosmological constant from a typical dust distribution ($\om \approx 0$), {\it i.e.} one sets $R_0 = 4\La$. From \eq{scaling}, this translates into the simple relationship $\la = \ep$.

The boundary of the spherical body corresponds to

\beqa \label{radiuscc} r_s &=& {d_H \over \sqrt{6\Om_\La}} \sqrt{ \la \over 1 + 2 \la } > 0 \rightarrow \\ \nonumber \la &>&0~~~~ \vee~~~~ \la < -1/2 ~~. \eeqa

\begin{figure} 

\epsfxsize=\columnwidth \epsffile{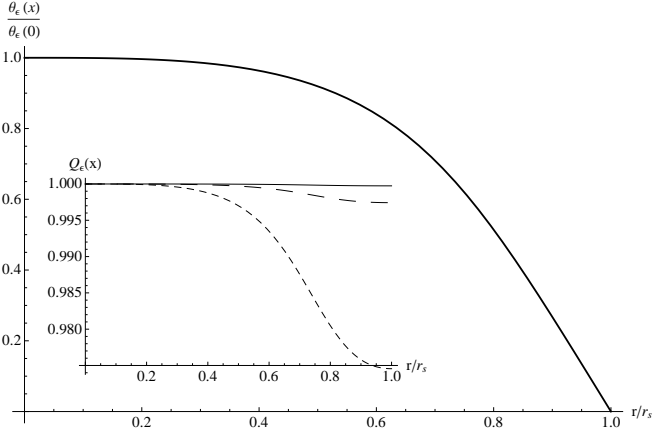}
\caption{Main: scaled solution $\th_0(x)/\th_0(0)$ for $\ep = \la = 0$. Inset: ratio $Q_\la$ for $\ep = 10^{-3}$ (full), $10^{-2}$ (dashed) and $10^{-1}$ (dotted).}
\label{thetan}

\end{figure}

The condition $r_s \ll d_H$ becomes

\beq {\la \over 1 + 2 \la }  \ll {6\Om_\La} \sim 1 \rightarrow \la =\ep \sim 0 ~~.  \eeq

\noindent so that \eq{difeqx} reads

\beq  \label{difeqxcc} \th'' - {2 \over x} \th' + {1 \over  x^2} \th (\th')^2 + {1 \over 6 x } (\th')^3 \left[ {1 \over x^2}(1-2\th^2) + 1 \right] = 0~~.\eeq

\noindent with boundary conditions $\th(1)=0$ and $\th'(1)= -\sqrt{6}$. Its (normalized) solution is plotted in Fig. \ref{thetan}.

Numerically, one finds that the solution for \eq{difeqx} with different values of $\la = \ep \sim 0$ (dubbed $\th_\la(x)$) are all very similar to the solution $\th_0(x)$ of \eq{difeqxcc} (where $\la = \ep =0$); this is visible in the inset of Fig. \ref{thetan}, which plots the normalized ratio between $\th_\la$ and $\th_0$,

\beq Q_\la \equiv {\th_\la (x) \over \th_0 (x)} {\th_0(0) \over \th_\la(0)} \sim 1~~. \eeq
 
All obtained solutions exhibit a smooth transition from $|\th_\la(0)| \sim 1$ to $\th_\la(1)=0$, and $\th_\la'(0)=0$. This good behaviour is illustrated by the central value, given by $\th_\la(0) \approx 0.9$ for all $\la = \ep \leq 0.1$. This similarity enables one to fit them through
	
\beq {\th_\la(x) \over \th_\la(0)} = 1 + 0.38 x^2 - 1.50 x^3 + 0.09 x^4 ~~, \label{fit} \eeq

\noindent with the weak dependency

\beq {\th_\la(0) \over 0.907} = 1 + 7.83 \times 10^{-3} \la + 5.95 \times 10^{-2} \la^2 ~~.\eeq

\noindent Notice, however, that the above fitting is merely illustrative, as it does not fulfill the required boundary conditions.
 
One may now set the constant $K$,

\beq K = -{\rho_c \over \th_\la(0) \ep}\sqrt{ 1 + 2 \la \over 2 } \approx -{\rho_c \over \th_\la(0) \la}\sqrt{ 1 \over 2 } \approx -0.78 {\rho_c \over \la} ~~.\eeq

Since \eq{radiuscc} may be approximated by

\beq \label{rsmimic} r_s \approx {d_H \over \sqrt{6\Om_\La}}\sqrt{ \la } \rightarrow \la \approx 6\Om_\La \left({r_s \over d_H}\right)^2~~,  \eeq

\noindent one may rewrite the above as

\beq K \approx -0.19 \left({ d_H \over r_s}\right)^2 \rho_c~~.\eeq

\noindent The first condition of \eq{rsmimic} is illustrated in Table I.



\begin{table}
	\begin{tabular}{c c}
		$\log \ep$		& $r_s$ \\
		\hline
		$-1$&$ 0.57~Gpc  $\\
		$-2  $&$ 1.97~Gpc  $\\
		$-10 $&$ 19.8~kpc  $\\
		$-20 $&$ 0.20~pc  $\\
		$-30 $&$ 0.41 ~AU  $\\
		$-40 $&$ 612~km  $\\
		$-50  $&$ 6.12~m $
	\label{table}
	\end{tabular}
	\caption{Table with radius of spherical distribution for different values of $\la = \ep \sim 0$.}
\end{table}


\subsection{$M = M_G$ condition}

Recalling that the condition $M = M_G$ implies a specific radius for the dust distribution, one writes

\beq r_s = 2\sqrt{\ka \th_\la(0) \over 3 \rho_c I} \sim \sqrt{\rho_H \over \rho_c} 11.4~\mpc ~~, \eeq

\noindent having used the solution $\th(x)$ with $\ep = \la = 0$ ({\it i.e.} $\th_0(x)$ instead of $\th_\la(x)$), for simplicity, in the numerical integration of $I$, and  $\rho_H = 1~ \mathrm{H ~atom / cm^3} = 1.660 \times 10^{-21} ~ kg / m^3$.

The typical distance between galaxies is of the order of  $r_s \sim 1 \mpc$ ({\it e.g.} Andromeda is $778~\mathrm{Kpc}$ away); one thus obtains

\beq \rho_c \approx 130 ~\mathrm{H ~atom / cm^3} ~~,\eeq

\noindent close to the lower bound of $H II$ star-forming regions. The interstellar medium attains densities up to $10^4 ~\mathrm{H ~atom / cm^3}$, allowing for a radius of the constant curvature bubble down to $\sim 100~\mathrm{Kpc}$.

In the $\la = \ep \sim 0$ regime, \eq{rscondition} becomes

\beqa \label{rsconditioncc} \la &=& \ep = {\Om_\La  \over 2\pi } {H_0^2\over G\rho_c}{\th(0) \over I } = 3.05 \times 10^{-5} {\rho_H \over \rho_c} \rightarrow \\ \nonumber \rho_c &>& 3.05 \times 10^{-5}~ \mathrm{H ~atom / cm^3} ~~, \eeqa

\noindent yielding a lower bound for $\rho_c$ well below the typical minimum density of the interstellar medium.

\subsection{Metric elements}

With $\la = \ep \sim 0$, the metric elements read

\beqa \label{metricscalingcc} e^{2\si(x)} &\approx& { 1\over 2 }\left[ { \th'(x) \over   x}\right]^2 ~~,\\ \nonumber e^{2\nu(x)} &\approx& {x^2 \over 3} \exp \left[ 2\int_x^1 {1 \over x^2}\th(x) \th'(x) dx \right]~~.\eeqa

One may compare this with the usual metric derived from GR for an interior solution, assuming for simplicity a vanishing pressure dust distribution, $p = 0$:

\beqa \label{interiorGRmetric}m_G(r) &\equiv & 4\pi \int_0^r \rho(r) r^2 dr ~~, \\ \nonumber e^{2\si} &=& \left( 1 - {m_G(r) \over 8\pi \ka r}\right)^{-1}~~, \\ \nonumber e^{2\nu} &=&\exp \left[ - \int_0^r {m_G(r) \over r\left( 8\pi\ka r- m_G(r)\right)} dr \right]~~, \eeqa

\noindent so that $M = M_G = m_G(r_s)$. In terms of the rescaled quantities given by \eq{scaling}, one reads

\beqa \label{interiorGRmetricscaling }m_G(x) &=& {M \over I} \int_0^x \th(x) x^2 dx~~, \\ \nonumber e^{-2\si} &=& 1- {2 \over 3}{m_G(x) \over M x}~~, \\ \nonumber e^{2\nu} &=&\exp \left[ - \int_0^x {m_G(x) \over x \left({3 \over 2}M x - m_G(x)\right)} dx \right]~~, \eeqa

\noindent having used $M=(16\pi / 3) \ka r_s $.

It is advantageous to evaluate analytically the above integrals; for this, one resorts to the quartic cubic fit of the $\ep = \la = 0$ solution \eq{fit}. One finds that the obtained metric coefficients are rather different from their GR counterparts (matching only at $x = 1$, due to the condition $M=m_g(x=1)$. However, this is to be expected: our solution arises from $R = R_0 = \mathrm{const.}$, while in GR one has $R = -T = \rho$.

\begin{figure} 

\epsfxsize=\columnwidth \epsffile{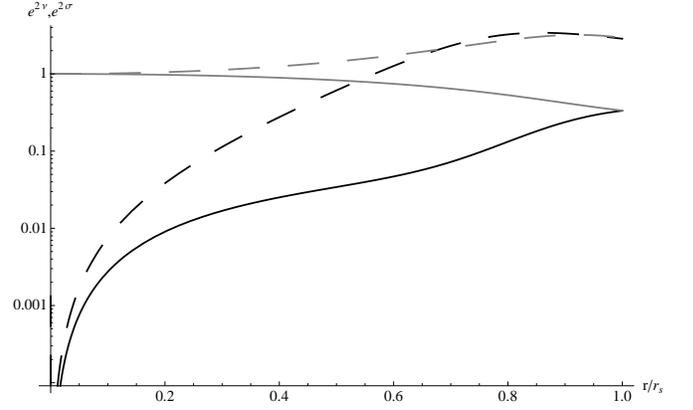}
\caption{Comparison of the metric coefficients $e^{2\nu}$ (full) and $e^{2\si}$ (dashed) derived from the $\ep = 0$ solution (black) and the usual GR interior solution (grey).}
\label{comparison}

\end{figure}

\section{Specific models}

One now looks at specific models that enable the mimicking of a cosmological constant, {\it i.e.}

\beq \ep = \la = {8 \over 1+\om}{F_2 \La \over f_2} \sim 0 \label{epdef} \eeq

Instead of obtaining the value for $\ep$, one discusses the range of model parameters derived from the condition $\ep \sim 0$ for several forms of $f_2(R)$. One sets $\om = 0 $, typical of a pressureless dust distribution; however, the above shows that the effect of a non-vanishing EOS parameter does not change the order of magnitude of the results.

\subsection{Power-law coupling}

One first assumes 

\beq f_2(R) = 1 + \left({R \over R_n}\right)^n~~, \eeq

\noindent so that

\beqa \label{power} \ep &=& 2 {F_2 R_0 \over f_2 } = 2n \left[ 1 + \left({R_n \over R_0}\right)^n \right]^{-1} \sim 0 \rightarrow \\ \nonumber n & \sim & 0 ~~~~ \vee ~~~~ \left({R_n \over R_0}\right)^n \gg 1  ~~. \eeqa

If $n > 0$, one obtains $R_n \gg R_0$; negative exponents imply $R_n \ll R_0$. In the following, one explores specific values for the exponent $n$.

\subsubsection{Linear coupling}

Replacing $n=1$, one must take into account the constraint arising from the Starobinsky scenario for inflation and the ensuing reheating process due to the presence of a linear coupling between curvature and matter, as described in Ref. \cite{reheating},

\beq R_1 = {9 \over 2\xi} (10^{-6}M_P)^2 = {4.4 \over \xi} \times 10^{56} ~m^{-2}~~,\eeq

\noindent with $1 < \xi < 10^4$. Since $R_1 \gg R_0 = 4\La$, and recalling \eq{cc}, one gets

\beq \ep = 8{\La \over R_1} =  {24 \Om_\La \over d_H^2 R_1} = 2.2 \xi \Om_\La \times 10^{-108} ~~, \eeq

\noindent which is vanishingly small, as intended, for any value of $\xi$ in the considered range.

\subsubsection{Quadratic coupling} 

One now aims at obtaining spherical mass distributions of constant curvature arising from a quadratic correction $n=2$; one imposes that this correction is below the linear one previously addressed at the inflationary stage, when $R \sim R_I \leq (M_P c/h)^2$,

\beqa \left({R_I \over R_2}\right)^2 &\ll& {2\xi \over 9} {R_I \over M_P^2} \times 10^{12} \rightarrow \\ \nonumber R_2 &\gg & {3 \over \sqrt{2\xi}} M_P^2 \times 10^{-6} \sim {2\over \sqrt{\xi}} \times 10^{62} ~m^{-2} ~~. \eeqa

From \eq{epdef}, one has

\beq \ep = 64 \left({\La \over R_2}\right)^2 = 576 \left({ \Om_\La \over d_H^2 R_2}\right)^2 \ll 2.3\xi \times 10^{-227}~~,\eeq

\noindent showing that any subdominant quadratic coupling at inflation allows for the local mimicking of the cosmological constant.

\subsubsection{Inverse power-law coupling}

Taking $n=-1$, dark matter constraints require that $R_{-1} \sim (10~Gpc)^{-2}$ \cite{DM}. Thus, one obtains from \eq{epdef}

\beq \ep = -2 {R_{-1} \over R_0} = - { 1 \over 6 \Om_\La}\left({d_H \over 10~Gpc}\right)^2 = 4.2 \times 10^{-2} ~~. \eeq

Analogously, setting $n=-1/3$ implies $R_{-1/3} \sim (10^6~Gpc)^{-2}$ \cite{DM}, leading to

\beqa \ep &=& - {2 \over 3} \left({R_{-1/3} \over R_0}\right)^{1/3} = {2 \over 3} \left({R_{-1/3} d_H^2 \over 12 \Om_\La }\right)^{1/3} = \\ \nonumber &=& {2 \over 3} {1 \over \left(12 \Om_\La \right)^{1/3}} \left({ d_H \over 10^6~Gpc }\right)^{2/3} \sim 8.6 \times 10^{-5} ~~. \eeqa

Both values for $\ep$ are vanishingly small, showing that a inverse power-law coupling between matter and curvature can account for the rotation curves of galaxies and enable the description of spherical bodies with constant curvature $R_0 = 4\La$.

\subsection{Exponential coupling}

One now assumes a power-law exponential coupling of the form

\beq f_2(R) = \exp \left[ \left({R \over R_n}\right)^n \right] ~~, \eeq

\noindent so that

\beqa \ep = 2{F_2 R_0 \over f_2} = 2n \left({R_0 \over R_n}\right)^n ~~. \eeqa

\noindent This expression is identical to \eq{power}, and the conclusions apply, namely that either $n \sim 0$, $R_n \gg R_0$ (positive $n$) or $R_n \ll R_0$ (negative $n$).

\section{Conclusions and Outlook}

In this work one shows that the rich phenomenology of non-minimally curvature-matter coupled models allows for a striking result, namely the possibility of inhomogenenous spherical mass distributions with constant curvature. The density profile of such matter distribution was obtained under general assumptions, and shown to lead to physically admissible values for density and size, typical of the interstellar environment. Finally, one establishes the compatibility between the matter distribution here discussed and the previously obtained constraints on the parameters of a class of power-law or exponential couplings $f_2(R)$. Although the mimicking of a CC is, by construction, local, one tentatively argues that the averaging of several of the obtained matter distributions could lead to a universal CC, which hints a novel mechanism for the mimicking of a CC type term.

\end{document}